\documentclass[12pt]{article}
\usepackage[centertags]{amsmath}
\usepackage{amsfonts}
\usepackage{amssymb}
\usepackage{amsthm} 
\usepackage{newlfont}
\usepackage{epsfig}
\usepackage{amscd}
\usepackage{graphicx}
\usepackage{epsfig}
\usepackage{amscd}

\newcommand{\beq}{\begin{equation}}
\newcommand{\eeq}{\end{equation}}
\newcommand{\ba}{\begin{array}}
\newcommand{\ea}{\end{array}}
\newcommand{\bea}{\begin{eqnarray}}
\newcommand{\eea}{\end{eqnarray}}

\usepackage{color}

\begin{document}

\begin{center}
{\large \sc \bf { Information propagation  in a quantum system. Examples of open spin-1/2 chains}
}

\vskip 15pt

{\large 
 A.I.~Zenchuk 
}

\vskip 8pt

{\it Institute of Problems of Chemical Physics, RAS,
Chernogolovka, Moscow reg., 142432, Russia, e-mail:   zenchuk@itp.ac.ru
 } 
\end{center}


\begin{abstract}
If the information  is encoded into the state of the  subsystem $S$  of a quantum system initially (at $t=0$), then it  becomes distributed  over the whole quantum  system at  $t>0$ due to the quantum interactions. Consequently, this information, in general, can be extracted, either completely or partially, from any  subsystem of a quantum system. {We suggest a method  of extraction of information, which 
is based on the  polarization measurements on the receiver  $R$}.
\end{abstract}

\section{Introduction}
\label{Sec:Introduction}

The problem of quantum information transfer 
is directly related to the construction of quantum communication systems. 
In  particular, this problem  was reduced to the problem of quantum state transfer,
which was first formulated by Bose in ref. \cite{Bose} for the Heisenberg homogeneous chain of spin-1/2 particles. However, the effect of the state transfer  was observed before Bose  in experiments with so called Loshmidt echo \cite{CPW,FBE}.
Many different modifications of the state transfer process  have appeared after that. They are  aimed at the improving of the state transfer characteristics, such as the fidelity \cite{Bose} and the  state transfer distance.
For instance, the inhomogeneous spin-1/2 chains with special values of coupling constants  \cite{CDEL,ACDE,KS,KF,KZ,GKMT}, the information flux approach \cite{FPPK,FPK} and the spin-1/2 chain in inhomogeneous magnetic field 
\cite{BGB,DZ} have been proposed for this purpose. In refs. \cite{BB,YBB} the entanglement has been
used as a resource which increases the fidelity and speeds up  the information transfer.  The perfect state transfer with the time dependent Hamiltonian  is proposed in \cite{FPK}.

However, all suggested methods of the perfect (or high probability)  state transfer have two basic obstacles for realization: (i)
the parameters of a quantum system (such as the coupling constants or/and the local magnetic field) must be fine tuned  which is a complicated task for the experimental realization and (ii)
the state transfer time interval  must be  fixed  and a minor deviation from its value
 reduces the fidelity of the state transfer.

At the same time,  the requirement to transfer the state of a qubit (or, more generally, the state of a subsystem)  $S$  to the qubit (subsystem) $R$  might be too rigorous.  Recall that  by the state of subsystem $S$ (or $R$) we mean the density matrix  $\rho^S(t)$ (or $\rho^R(t)$) redused with respect to all  nodes of a quantum system, except nodes of the subsystem $S$ ( or $R$). Here the parameter $t$ is the usual time. Therefore,  we know all information encoded into the subsystem $S$ if we know all elements of the matrix $\rho^S$. We say that the information encoded into  the initial state of the subsystem $S$ (i.e. in the matrix $\rho^S(0)$) is completely transfered to the subsystem $R$  at the instant $t=t_1$ if we may obtain all elements of $\rho^S(0)$ performing some operations (either quantum or classical) on the elements of $\rho^R(t_1)$. 

Therefore, in order to transfer the information encoded into  the state of the subsystem $S$, it is not necessary to transfer the state itself. 
The  information transfer is much simpler for the experimental realization in comparison with the state transfer, { which will be demonstrated below in this paper}.  In fact,
 the information encoded into the  original state of some subsystem $S$ of a spin system  spreads over the whole system 
due to the quantum interactions among all nodes.
Thus, in principle,  it can be obtained (either partially or completely) from the analysis of the state of any other subsystem of a quantum system. This is a general property of a quantum system governed by any Hamiltonian, { and namely this property  will be used hereafter}. Due to this property we do not need to  adjust precisely parameters of a communication system (such as the coupling constants and the local magnetic field) { in oder} to achieve the complete information transfer between two subsystem. Moreover, the information can be extracted at (almost) any time moment $t_1>0$. We see that the information transfer is not subjected to the two above obstacles for the practical realization, which appear in the realization of the state transfer.  
 A simple model of the information transfer  through the two qubit system  $B\cup A$  supplemented by a one qubit sender $C$ is considered in \cite{WHDS}. The information was transfered  from the qubit $C$ to the qubit $A$ (receiver) due to  the projective measurement over the two qubit  subsystem $C\cup B$.  In our paper we provide the information transfer from the sender to the receiver  by  means of the evolution operator 
acting on the whole system including the sender.

However, even if the state of  the receiver $R$  at the instant $t_1$ contains the complete  information encoded into the initial  state of the sender $S$, there is a problem of extracting of this information from  $R$, i.e. the problem of defining the elements of the matrix $\rho^R(t_1)$ in practice. This  problem is studied in the quantum tomography  \cite{JKMW}.
 We propose a measurement based method of the density matrix reconstruction \cite{H,BAPS,HSBR,RHJ}, where the multi-channel communication system is used in combination  with the appropriate  polarization measurements at the end of each channel.

One has to emphasize that the  extraction of information by means of measurements is not necessary,  for instance, in the case of the data transfer  in the quantum computer { during the computational process}. In fact, let the data be encoded into the elements of $\rho^S(0)$ and the matrix 
$\rho^R(t_1)$ is { linearly} related with the elements of the matrix $\rho^S(0)$: $\rho^R(t)=L_t(\rho^S(0))$. Then the computation algorithm must involve the algorithm solving the above  relation with respect to the elements of $\rho^S(0)$. This is a usual mathematical problem which must be resolvable by the quantum computation algorithms \cite{NC}.

This paper is organized as follows.
The description of the information transfer  algorithm  is represented in Sec.\ref{Section:inf}. Examples of the one-qubit density matrix information transfer along the { open}  spin chains  are given in Sec.\ref{Section:examples}. { A possible three-channel scheme of the experimental realization of the one-qubit information  transfer } is represented in Sec.\ref{Section:exp}. Conclusions are given in Sec.\ref{Section:conclusions}.

\section{Evolution of information encoded into initial quantum state}
\label{Section:inf}
\label{Section:ev}

\begin{figure*}
   \epsfig{file=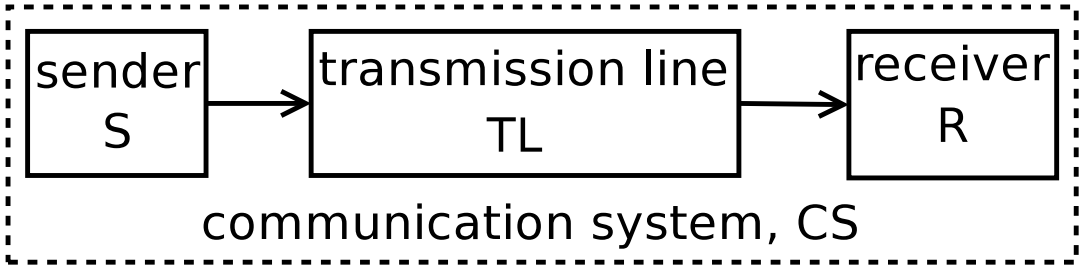
   ,scale=0.65,angle=0}
\caption{General  scheme of the quantum one-channel communication system}
 \label{Fig:S_TL_R}
\end{figure*}

Let $CS$ (communication system) be a quantum system of $N$ spin-1/2 particles {
(these particles will be called nodes below in this article, so that the above system is  a spin system of $N$ nodes)}.
Let $S$ (sender)  and $R$ (receiver) be two different  subsystems of the system $CS$. We denote by $TL$ (transmission line)
the rest of the quantum system, so that $CS=S\cup TL \cup R$, see Fig.\ref{Fig:S_TL_R}. Let the subsystems $S$, $TL$ and $R$
consist of respectively $N_S$, $N_{TL}$ and $N_R$ nodes with $N_S+N_{TL}+N_R =N$.
We associate the Hilbert spaces $H_S$, $H_{TL}$ and $H_R$ with respectively $S$, $TL$ and $R$.

The state of the { whole system $CS$ } is  described by the density matrix $ \rho$ with elements 
$\rho_{\alpha;\beta}$, where $\alpha=(\alpha_1,\dots,\alpha_N)$ and $\beta=(\beta_1,\dots,\beta_N)$ are  multi-indices, $\alpha_i,\beta_i=0,1$, while the states of the subsystems $S$, $TL$ and $R$ are described by the proper reduced matrices:
\begin{eqnarray}
\rho^S={\mbox{Tr}}_{TL,R} \rho\in H_S,\;\;
\rho^R={\mbox{Tr}}_{S,TL} \rho\in H_{R},\;\;
\rho^{TL}={\mbox{Tr}}_{S,R} \rho \in H_{TL}.
\end{eqnarray}
The obvious necessary condition for the complete information transfer is $\dim H_S \le \dim H_R$, because namely this case guarantees that the number of free elements in the matrix $\rho^S$ does not exceed { the number of free elements } in  the matrix $\rho^R$. 
Let the evolution of the system be governed by some Hamiltonian $H(t)$. Then the spin dynamics is described by the density matrix $\rho(t)$, which is 
  the solution to the Liouville equation ($\hbar=1$),
\begin{eqnarray}
i\frac{\partial \rho(t)}{\partial t} = [H(t),\rho(t)].
\end{eqnarray}
{This solution  reads}:
\begin{eqnarray}\label{rhot}
 \rho(t)= U(t)  \rho(0)  U^+(t),\;\;\; U(t)=\exp(-i \int^t H(t') dt'),
\end{eqnarray}
where $ \rho(0)$ is the initial density  matrix.

We say that the information, which is  initially (at $t=0$) encoded into the state of  $S$ (i.e. all elements of the   initial density matrix $\rho^S(0)$), is completely (partially) transfered  to the subsystem $R$ at the instant  $t=t_1$ if it may be completely (partially) extracted  from the analysis of the density matrix $\rho^R(t_1)$ at instant $t_1$. It is obvious that the 
information transfer requires some
 relations between the elements of the density matrices $\rho^R(t_1)$ and $\rho^S(0)$. { Eq.(\ref{rhot}) means that} 
 these relations are  linear, and they may be schematically written { as the following} matrix equation:
\begin{eqnarray}
\rho^R(t_1)=L_{t_1} (\rho^S(t_0)),
\end{eqnarray}
where $L_{t_1}$ is a linear operator depending on a particular instant $t_1$.
Assume the existence of such operator $L_{t_1}^{-1}$ that
\begin{eqnarray}
\rho^S(0)=L_{t_1}^{-1}(\rho^R(t_1)).
\end{eqnarray}
If $L_{t_1}^{-1}$ may be uniquely constructed, then the information is completely transfered from $S$ to $R$ at the instant $t=t_1$, i.e. all elements of $\rho^S(0)$ can be uniquely constructed. 
If $L_{t_1}^{-1}$ is not unique, then the information is partially transfered from $S$ to $R$
(we may not uniquely find all elements of the matrix $\rho^S(0)$ in this case). If $L_{t_1}^{-1}$ does not exists, then the information may  not be transfered at the instant $t_1$. 

{ Thus we relate the possibility to completely (partially) transfer the information from the sender $S$ to the receiver $R$  with the condition of  unique (non-unique) resolvability of the linear system (\ref{rhot}) with respect to the elements of the  initial  density matrix of the sender $\rho^S(t_0)$, which is a well known condition of the linear algebra.}
{
Note, that the perfect state transfer from the sender $S$ to receiver $R$ \cite{CDEL,ACDE,KS} corresponds to the very special form of the linear operator $L_{t_1}$, namely $L_{t_1} (\rho^S(t_0)) \equiv \rho^S(t_0)$, which has been used in the above references as well as  in other references concerning either the ideal  or the high probability state transfer \cite{KF,KZ,GKMT,FPPK,FPK,BGB,DZ} (note that the density matrix $\rho^S(t_0)$ is not completely arbitrary in those references because it is associated with the pure state of the one-node sender; the sender and the receiver are respectively  the first and the last nodes of the spin chains). Of course, the condition for the unique resolvability of the system (\ref{rhot}) with respect to the elements of the density matrix  $\rho^S(t_0)$ is valid because  $\rho^R(t_1)\equiv \rho^S(t_0)$ in this case.  But this condition may be valid even if the condition for the perfect state transfer is not verified. For this  reason we may conclude that the condition for the perfect state transfer is embedded into the condition for the complete information transfer.  As a consequence, the information transfer may be organized in a much simpler way then the ideal state transfer.  To give more details  regarding this conclusion, we consider a particular example   of  the communication system, where 
 both the sender $S$ and  the receiver  $R$ consist of a single  spin-1/2 particle.}
  Generalization to the sender (receiver) consisting of more number of particles is straightforward. 

Let the first and the $N$th nodes be sender $S$ and  receiver $R$ respectively, i.e.
\begin{eqnarray}
\rho^S= \{\rho^S_{\alpha_1;\beta_1}\},\;\; 
{ \rho^{TL} = \{ \rho^{TL}_{\alpha_2\dots \alpha_{N-1};\beta_2\dots \beta_{N-1}}\},}
\;\;\rho^R= \{\rho^R_{\alpha_N;\beta_N}\}\;\;\;\alpha_i,\beta_i=0,1.
\end{eqnarray}
{ We derive the condition, for which  the elements of $ \rho^S(t_0)$ may be found from the known elements of the reduced density matrix $\rho^R(t_1)$ at some instant  $t_1>0$. Then  we demonstrate that this condition is much less restrictive then the condition   for the ideal transfer of the initial state $\rho^S(t_0)$ to the receiver $R$ at some instant $t_1$. }

First, we write
\begin{eqnarray}
\rho_{\gamma;\delta}(t)=\sum_{\alpha,\beta} U_{\gamma;\alpha}(t)\rho_{\alpha;\beta}(0) U^+_{\beta;\delta}(t)=
\sum_{\alpha_1,\beta_!} T_{\gamma\delta; \alpha_1\beta_1}(t) \rho^S_{\alpha_1;\beta_1}(0),
\end{eqnarray}
where 
\begin{eqnarray}
T_{\gamma\delta; \alpha_1\beta_1}(t)=\sum_{\alpha_1,\dots,\alpha_N,\beta_2,\dots,\beta_N} U_{\gamma;\alpha}(t)
\tilde \rho_{\alpha_2\dots \alpha_N;\beta_2\dots \beta_N}(0) U^+_{\beta;\delta}(t),
\end{eqnarray}
and { $\tilde \rho_{\alpha_2\dots \alpha_N;\beta_2\dots \beta_N}$ are the elements of the matrix $\tilde \rho$, which is the matrix $\rho$ reduced with respect to the sender $S$: 
$\tilde \rho = {\mbox{Tr}}_S \rho = \{\tilde \rho_{\alpha_2\dots \alpha_N;\beta_2\dots \beta_N}\}\in H_{TL}\cup H_R$.}
The reduced density matrix $\rho^R(t)$  reads
\begin{eqnarray}\label{rho_red2}
\rho^R_{\gamma_N;\delta_N}(t)= \sum_{\alpha_1,\beta_1} T^R_{\gamma_N\delta_N;\alpha_1\beta_1}(t) \rho^S_{\alpha_1;\beta_1}(0) ,
\end{eqnarray}
where
\begin{eqnarray}\label{T}
&&
T^R_{\gamma_N\delta_N;\alpha_1\beta_1}(t)= \sum_{\gamma_1,\dots,\gamma_{N-1}}
 T_{\gamma_1\dots\gamma_{N-1}\gamma_N\gamma_1\dots\gamma_{N-1}\delta_N;\alpha_1\beta_1}(t)=\\\nonumber
&&
\sum_{{\gamma_1,\dots,\gamma_{N-1},\alpha_2,\dots,\alpha_{N-1}} \atop{
\beta_2,\dots,\beta_{N-1}}} U_{\gamma_1,\dots,\gamma_{N-1}\gamma_N;\alpha}(t)\tilde \rho_{\alpha_2\dots\alpha_N;\beta_2\dots\beta_N}(0) U^+_{\beta;\gamma_1,\dots,\gamma_{N-1}\delta_N}(t).
\end{eqnarray}
The information about $\rho^S(0)$ may be completely extracted from $\rho^R(t_1) $ at some instant $t=t_1$ if the system (\ref{rho_red2}) may be uniquely solved with respect to the elements of the density matrix $\rho^S(0)$ at the instant $t_1$.  { In other words, one has single scalar condition
\begin{eqnarray}\label{IT}
\label{Det}
&&
{ \det \;T^R(t_1)\neq 0,}\\\label{TR}
&&
T^R(t)=\left(
\begin{array}{cccc}
T^R_{00;00} (t) & T^R_{00;01} (t) & T^R_{00;10} (t) &T^R_{00;11} (t) \cr
T^R_{01;00} (t)  & T^R_{01;01}(t)  & T^R_{01;10} (t) &T^R_{01;11}(t)  \cr
T^R_{10;00} (t)  & T^R_{10;01}(t)  & T^R_{10;10}(t)  &T^R_{10;11}(t)  \cr
T^R_{11;00} (t)  & T^R_{11;01}(t)  & T^R_{11;10}(t)  &T^R_{11;11} (t) 
\end{array}
\right) .
\end{eqnarray}
On the other hand, the conditions for the ideal state transfer at the instants $ \tau_i$ read
\begin{eqnarray}\label{PST}
T^R_{\gamma_N\delta_N;\alpha_1\beta_1}(\tau_i) =\delta_{\gamma_N\alpha_1} \delta_{\delta_N\beta_1},\;\;
\gamma_N,\delta_N,\alpha_1,\beta_1 = 0,1,
\end{eqnarray}
{ which is a system of $2^4=16$ scalar equations. Namely in this case eq.(\ref{rho_red2}) reduces to the identity  $\rho^R(\tau_i)\equiv  \rho^S(0)$. Emphasize that the condition for the complete information transfer (\ref{IT}) is valid in this case, because the matrix $T^R(\tau_i)$ in formula (\ref{TR}) becomes the $4\times 4$  unit matrix: $T^R(\tau_i)=I_4$.} { Therefore, condition (\ref{PST}) is a {\it very particular realization}  of  condition (\ref{Det}).   }  Thus we conclude that the  conditions of the perfect state transfer (i.e. system(\ref{PST})) are embedded in the condition of the complete information transfer (\ref{IT}).  
{ Eq.(\ref{IT}) states that the complete information transfer is impossible only at such time moments  $t=\tilde \tau_i$ that 
\begin{eqnarray}\label{Det_0}
 \det \;T^R(\tilde \tau_i) = 0.
\end{eqnarray}
}
{ This remark suggests us to formulate the}  principal difference between conditions (\ref{IT})  and (\ref{PST})  as follows: there 
is a set of  instants  $\tilde \tau_i$ (solutions to eq.(\ref{Det_0})), when the complete information transfer {\it is impossible},
whereas  there is  another set of  instants $ \tau_i$ (solutions to the conditions (\ref{PST})) when the 
perfect state transfer {\it is possible}. 
This simple and straightforward  analysis allows one to conclude that the complete information transfer may be organized much simpler than the perfect state transfer. If the condition (\ref{IT}) is not satisfied, than the rank of the matrix $T^R$ in  formula (\ref{IT}) is less then four, so that only part of information about the density matrix $\rho^S(t_0)$ may be obtained from the analysis of the density matrix $\rho^R(t_1)$.}

{ Now we consider the information transfer in more details and} give a different form to the condition (\ref{IT}) for the unique resolvability of the system (\ref{rho_red2}). First of all, we remark 
that there are only three arbitrary real parameters $x_i$, $i=1,2,3$, in the original one-qubit matrix $\rho^S(0)$, which parametrize all elements of $\rho^S(0)$ as follows (hereafter $\rho_{ij}\equiv \rho_{i;j}$, $i,j=0,1$):
\begin{eqnarray}
&&\rho^S_{00}(0)=x_1,\;\;\;\rho^S_{01}(0)=x_2+i x_3,\;\;\;\rho^S_{10}(0)=x_2-i x_3,\;\;\;\rho^S_{11}(0)=1-x_1.
.
\end{eqnarray}
Being the density matrix,  $\rho^S$ must be a positive-semidefinite matrix, so that the  values of the parameters 
$x_i$ ($i=1,2,3$) are restricted by the condition $(1-2 x_1)^2 +(2 x_2)^2 +(2 x_3)^2 \le 1$.
Taking into account that
$\rho^R_{11}=1-\rho^R_{00}$ and $\rho^R_{01}=(\rho^R_{10})^*$ 
(were the star means the complex conjugate value),
we reduce  the system of four linear equations  (\ref{rho_red2})  
to the  equivalent system of three linear equations for  the parameters $x_i$, $i=1,2,3$:
\begin{eqnarray}\label{A}
Re(\rho^R_{01}(t_1))&=&\sum_{i=1}^3  A_{1i}(t_1) x_i +A_{10}(t_1),\;\;\\\nonumber
Im(\rho^R_{01}(t_1))&=&\sum_{i=1}^3  A_{2i}(t_1) x_i +A_{20}(t_1),\;\;\\\nonumber
\rho^R_{00}(t_1)&=&   \sum_{i=1}^3  A_{3i}(t_1) x_i +A_{30}(t_1).
\end{eqnarray}
Here
\begin{eqnarray}\label{AT}
A_{11}&=&{\mbox{Re}}(T^R_{01;11}-T^R_{01;11}),\;\;A_{12}={\mbox{Re}}(T^R_{01;01}+T^R_{01;10}),\;\;A_{13}=-{\mbox{Im}}(T^R_{01;01}-T^R_{01;10}),\;\;
\\\nonumber &&
A_{10}={\mbox{Re}}(T^R_{01;11}),\\\nonumber
A_{21}&=&{\mbox{Im}}(T^R_{01;00}-T^R_{01;11}),\;\;A_{22}={\mbox{Im}}(T^R_{01;01}+T^R_{01;10}),\;\;A_{23}={\mbox{Re}}(T^R_{01;01}-T^R_{01;10}),\;\;
\\\nonumber && 
A_{20}={\mbox{Im}}(T^R_{01;11}),\\\nonumber
A_{31}&=&T^R_{00;00}-T^R_{00;11},\;\;A_{32}=T^R_{00;01}+T^R_{00;10},\;\;A_{33}=i(T^R_{00;01}-T^R_{00;10}),\;\; A_{30}=T^R_{00;11}.\\\nonumber
\end{eqnarray}
We see that the   information about the initial state of sender (i.e. the elements of $\rho^S(0)$)
 may be completely extracted  from $\rho^R(t_1)$ if the system (\ref{A}) is uniquely solvable for $x_i$, $i=1,2,3$.
This unique resolvability  requires
\begin{eqnarray}\label{detA}
\det \{A_{ij}(t_1), i,j=1,2,3\}\neq 0.
\end{eqnarray}
If condition (\ref{detA}) is satisfied at some instant $t_1$ (which means that ${\mbox{rank}}  \{A_{ij}(t_1)\}=3$), 
then the complete information about $\rho^S(0)$ may be obtained from $\rho^R(t_1)$. Otherwise, if
$0< {\mbox{rank}}  \{A_{ij}(t_1)\}< 3$, then 
 we may obtain only partial information about $\rho^S(0)$.  If ${\mbox{rank}}  \{A_{ij}(t_1)\}=0$, then no information may be extracted from the subsystem $R$ at the instant $t_1$. The later is possible only if $\{A_{ij}\}\equiv 0$.

{ In addition,} one should remark that, in principle,  the instants in three equations (\ref{A}) may be different.

{ An important  component of the information transfer is the initial state $\rho(0)$. It seemed out that not any $\rho(0)$ allows the complete information transfer, which is shown in Sec.\ref{Section:N4} (see the text after eq.(\ref{H4}))}.
In particular, $\rho(0)$ might have the following form:
\begin{eqnarray}\label{dp}
 \rho(0) =\rho^S(0) \otimes \tilde  \rho(0), \;\;{\mbox{Tr}}( \rho^S)={\mbox{Tr}}(\tilde  \rho) = 1,
\end{eqnarray}
where $\rho^S\in H_S $  and $\tilde \rho\in H_{TL}\cup H_R$. 
This case corresponds, for instance, to  the initial state  described by the following wave function
\begin{eqnarray}\label{in}
|\Psi(0)\rangle = |\Psi^S (0)\rangle \otimes |\Psi^{TL\cup R} (0)\rangle,
\end{eqnarray}
where $|\Psi^S\rangle$ is the state of the subsystem $S$ and $|\Psi^{TL\cup R}\rangle$ is the state of the subsystem $TL\cup R$.
Thus
\begin{eqnarray}
\rho^S (0)=|\Psi^S (0)\rangle \langle\Psi^S (0) |,\;\;\tilde \rho(0)= |\Psi^{TL\cup R} (0)\rangle \langle \Psi^{TL\cup R} (0)|.
\end{eqnarray}
The initial state (\ref{in}) with the one-qubit wave function { $|\Psi^S (0)\rangle=|1\rangle$ and 
$ |\Psi^{TL\cup R} (0)\rangle = |0\dots 0\rangle $ } is used in the usual problem of the quantum state transfer along the spin-1/2 chain \cite{Bose,CDEL,ACDE,KS,KF,KZ,GKMT}.

\section{Examples: spin-1/2 open chains}
\label{Section:examples}
\label{Section:example1}

{ As is shown above, }
there are no  rigorous special requirements to the initial density matrix $\rho^{TL\cup R}(0)={\mbox{Tr}}_S \rho(0)$ as well as to the evolution operator $U(t)$ { in order to organize  complete information transfer, whereas these requirements are very rigorous in the case of  perfect state transfer}.
Nevertheless, it is important to know, which Hamiltonian   { is preferable
in the  process of information  transfer}
and what   the appropriate initial state of the subsystem  $TL\cup R$ might be used.
Considering examples of the information transfer along the short spin-1/2 chains, {we show that the 
XY-Hamiltonian is a proper one. Of course, all settings taken for  perfect state transfer may be applied for 
the organization of  complete information transfer. But we show that the complete information transfer may be 
realized even in the cases when the perfect state transfer is impossible, see example of the four-node 
homogeneous spin chain in Sec.\ref{Section:N4}.}

Let the  initial density matrix be representable in the form (\ref{dp}). 
We consider the open homogeneous spin-1/2 chain of $N$ nodes governed by the Hamiltonian 
$H_{XY}$ using the nearest neighbor interaction approximation:
\begin{eqnarray}\label{H}
H_{XY}= -\sum_{i=1}^{N-1} \frac{D}{2} (I^+_iI^-_{i+1} + I^-_iI^+_{i+1}),
\end{eqnarray}
where  $D$ is the coupling constant between the nearest neighbors, 
$I^\pm_i=I_{x;i} \pm i I_{y;i}$ and $I_{\alpha;i}$, $\alpha=x,y,z$, are the projection operators  of the total spin angular momentum. Thus, the evolution operator reads:
\begin{eqnarray}
U(t)=e^{-i H_{XY} t }.
\end{eqnarray}
Let us consider the initial state in the form (\ref{dp}) 
with an arbitrary initial  density matrix of sender $\rho^S(0)$ and two types of initial density matrix $\tilde \rho(0)$:
\begin{eqnarray}\label{rho01}
1.  && \tilde \rho(0) = {\mbox{diag}}(1,\underbrace{0,\dots,0}_{N-2}),
\\\label{rho02}
2. &&
\tilde \rho(0) = \frac{e^{-\beta \tilde H_{XY}}}{{\mbox{Tr}} \left(e^{-\beta \tilde H_{XY}}\right)},
\\ &&\label{tH}
\tilde H_{XY}= -\sum_{i=2}^{N-1} \frac{D}{2} (I^+_iI^-_{i+1} + I^-_iI^+_{i+1}) + \sum_{i=2}^N \omega_i I_{zi},
\end{eqnarray}
where $\omega_i$, $i=1,\dots,N$, are Larmor frequencies.
The initial state (\ref{rho01}) corresponds to the usual arbitrary quantum state transfer problem \cite{Bose}, while the initial state (\ref{rho02})
is the  thermal equilibrium state of the subsystem $TL\cup R$ with the Hamiltonian $\tilde H_{XY}$.
It will be noted in Sec.\ref{Section:N4} that not any set of  $\omega_i$ in  Hamiltonian (\ref{tH})  is suitable for the complete information transfer.

\subsection{Three node chain}
\label{Section:N3}

Let $N=3$, so that the $TL$ consists of a single node similar to $S$ and $R$.
The information can be completely transfered, for instance, if
\begin{eqnarray}\label{N3rho01}
\tilde \rho(0)={\mbox{diag}}(1,0,0,0)
\end{eqnarray}
or  
\begin{eqnarray}\label{N3rho02}
\tilde \rho(0)=\frac{e^{-\beta \tilde H_{XY}}}{{\mbox{Tr}} \left(e^{-\beta \tilde H_{XY}}\right)},
\end{eqnarray}
where
\begin{eqnarray}\label{H3}
\tilde H_{XY}= - \frac{D}{2} (I^+_2I^-_{3} + I^-_2I^+_{3}).
\end{eqnarray}
The evolution of the whole system is governed by
the Hamiltonian (\ref{H}) with $N=3$.

For instance, if $\tilde \rho(0)$ is given by Eq.(\ref{N3rho01}) we obtain
the following nonzero elements of $T^R$, see eq.(\ref{T})  (we use the standard basis of vectors  $|00\rangle$, $|01\rangle$, $|10\rangle$, $|11\rangle$):
\begin{eqnarray}\label{H3T}
T^R_{00;00}=1,\;\;T^R_{01;01}=T^R_{10;10}=r, \;\;T^R_{11;11}=r^2,\;\;T^R_{00;11}=1-r^2,\;\;\;r=-\sin^2\frac{t}{2\sqrt{2}}.
\end{eqnarray}
It is remarkable that at the instant $t_p$ such that 
$r^2(t_p)\equiv \sin^4\frac{t_p}{2\sqrt{2}}=1$
we have
\begin{eqnarray}
\rho^R(t_p) = U_1 \rho^S(0) U_1^+,\;\;U_1={\mbox{diag}}(i,-i).
\end{eqnarray}
Thus,  applying the local unitary transformation to the state of the receiver at the instant $t_p$ we obtain the initial state of the sender. In other words, we have the perfect state transfer  from the first to the last node of the three node spin chain. 
This result coincides with \cite{CDEL}. 
Of course, the perfect state transfer means that the information is completely transferred from the subsystem  $S$ to the subsystem  $R$.

However, the information about $\rho^S(0)$ may be extracted at different instants using the results of Sec.\ref{Section:ev}. The only requirement to the time moment is predicted by  condition (\ref{detA}), which reads in our case:
\begin{eqnarray}\label{A3}
\det\; \{A_{ij}(t)\}=r^4(t) =  \sin^8 \frac{t}{2\sqrt{2}}\neq 0.
\end{eqnarray}
Here $A_{ij}$ are defined in Eqs.(\ref{AT}) with $T_{ij;nm}$ given by expressions (\ref{H3T}).
Eq.(\ref{A3}) means that the information may not be completely transfered only at the discrete set of time moments.


\subsection{Four node chain}
\label{Section:N4}
Let $N=4$.
The information can be completely   transfered if, for instance,
\begin{eqnarray}\label{N4rho01}
\tilde \rho(0)={\mbox{diag}}(1,0,0,0,0,0,0,0),
\end{eqnarray}
or  
\begin{eqnarray}\label{N4rho02}
\frac{e^{-\beta \tilde H_{XY}}}{{\mbox{Tr}} \left(e^{-\beta \tilde H_{XY}}\right)},
\end{eqnarray}
where
\begin{eqnarray}\label{H4}
\tilde H_{XY}= - \frac{D}{2} (I^+_2I^-_{3} + I^-_2I^+_{3}+I^+_3I^-_{4} + I^-_3I^+_{4}) + \omega_4 I_{z4}.
\end{eqnarray}
Evolution of the whole system is governed by
the Hamiltonian (\ref{H}) with $N=4$.
It is important that $\omega^4\neq 0$, because otherwise  ${\mbox{rank}} \; \{A_{ij}\}=1$  and the information 
may not be completely transfered  to the last node   in this case.

In the case of the initial density matrix  $\tilde \rho(0)$  given by Eq.(\ref{N4rho01}), one has the following nonzero elements of $T^R$, see eq.(\ref{T}):
\begin{eqnarray}\label{H4T}
&&
T^R_{00;00}=1,\;\;T^R_{01;01}=-T^R_{10;10}=ir, \;\;T^R_{11;11}=r^2,\;\;T^R_{00;11}=1-r^2,\\\nonumber
&&
r=\frac{1}{5+\sqrt{5}} \left(
2 \sin\frac{(1+\sqrt{5})t }{4}  + (3+\sqrt{5}) \sin \frac{(1-\sqrt{5})t }{4}
\right).
\end{eqnarray}
The perfect state transfer is impossible in this case, which  agrees with \cite{CDEL}. 
However, the complete information about $\rho^S(0)$ may be extracted at some instant $t_1$ using the results of Sec.\ref{Section:ev}.
 Condition (\ref{detA}) reads
\begin{eqnarray}\label{A4}
\det\; \{A_{ij}(t_1)\} = r^4(t_1)\neq 0.
\end{eqnarray}
Here $A_{ij}$ are defined in Eqs.(\ref{AT}) with $T_{ij;nm}$ given by expressions (\ref{H4T}).
Eq.(\ref{A4}) means that the information, in principle, can be extracted almost at any time moment, except the discrete set of instants, given by condition (\ref{A4}).

\section{Measurement based extraction of information}
\label{Section:exp}
 
{
Let us compare the perfect state transfer with the complete information transfer from the information extraction point of view. One should underline that both are very similar. In fact, as is  mentioned in the Introduction, the extraction of information from the receiver $\rho^R$ is not  needed if this information is an ''intermediate'' one  which is supposed to be used as data for the subsequent quantum computations. 
Then no additional operation is required, because, in both cases, all information about $\rho^S(0)$ is represented in $\rho^R(t_1)$. However, the extraction of information is necessary in the case of the communication systems, when one might need to read this information from the receiver { and use in the classical devises hereafter}. 
The information may be read  by the quantum tomography tool. This holds for both cases (the  perfect state transfer and the complete information transfer). Only if we need to transfer a single parameter of the density matrix (say $x_1$, like it takes place in the perfect state transfer considered in refs. \cite{Bose,CDEL,ACDE,KS}) one can be satisfied with a simple measurement of polarization on the receiver of a communication system.}

Hereafter we represent the  method of information extraction from the receiver. 

We see that in order to find $x_i$ from the system (\ref{A}) one has to know elements of the reduced density matrix $\rho^R(t_1)$.
The problem of  reconstruction of its elements  is a quantum tomography problem \cite{JKMW}. We use a method for  
construction of the elements  of $\rho^R(t_1)$ based on the  polarization   measurements \cite{H,BAPS,HSBR,RHJ}.

Let us consider the multi-channel  communication system with the non-interacting channels whose number equal to the number of arbitrary parameters in the density matrix $\rho^S(0)$, i.e. three channels for the one-qubit
density matrix $\rho^S(0)$. Each channel is equivalent to that represented in Fig.\ref{Fig:S_TL_R}. All channels have the same initial state and the spin dynamics is governed by the same Hamiltonian in each channel. 
To find the elements of $\rho^S(0)$  we measure the average polarizations $J_n(t_1)$  on the receivers of all  channels ($n=1,2,3$) 
at some instant $t_1$ in some  directions ${\mathbf a}_n=(a_{n1},a_{n2},a_{n3})$ (here $\sum_i a_{ni}=1$, $n=1,2,3$, and 
 $a_{ni}$ are  positive parameters):
\begin{eqnarray}\label{IN}
J_n(t_1)
 = {\mbox{Tr}} (\rho^R (t_1){\mathbf I_N}\cdot {\mathbf a}_n),\;\;I_N=(I_{Nx},I_{Ny},I_{Nz}),
\end{eqnarray}
where
 $I_{N\alpha}\equiv \frac{1}{2}\sigma_\alpha$ ($\alpha=x,y,z$) is the projection of the $N$th spin (receiver) on the axis $\alpha$, $\sigma_\alpha$ are Pauli matrices. Substituting $I_{N\alpha}$ into Eq.(\ref{IN}) we transform this equation into the following one:
\begin{eqnarray}\label{J}
J_n(t_1)= a_{n1} {\mbox{Re}}\; \rho^R_{01} (t_1)-  a_{n2} {\mbox{Im}}\; \rho^R_{01} (t_1)+a_{n3}\left( \rho^R_{00}(t_1)-\frac{1}{2}\right),\;\;n=1,2,3.
\end{eqnarray}
We see that the average polarization $J_n(t_1)$ contains the  information about all elements of $\rho^R(t_1)$. Similar to the matrix $\rho^S(0)$, all elements of the matrix $\rho^R(t_1)$ may be written in terms of three real  parameters: $\rho^R_{00}(t_1)$, ${\mbox{Re}} \rho^R_{01}(t_1)$ and ${\mbox{Im}} \rho^R_{01}(t_1)$. 
Therefore, if one can uniquely solve the system (\ref{J}) with respect to these parameters, then we will be able to express all elements of the density matrix $\rho^R(t_1)$ in terms of the polarizations $J_n(t_1)$, $n=1,2,3$.  The unique resolvability condition for the system (\ref{J}) reads
\begin{eqnarray}
\label{ani}
\det \{a_{ni}: n,i=1,2,3\}\neq 0.
\end{eqnarray}
Finally, we derive the relation between $x_i$ (parameters of the density matrix $\rho^S(0)$) and polarizations  $J_n(t_1)$. 
For this purpose, we 
eliminate elements of the density matrix $\rho^R(t_1)$ from the rhs of eq.(\ref{J})  using eqs.(\ref{A}):
\begin{eqnarray}\label{JJ}\label{JJ2}
J_n(t_1)= \sum_{i=1}^3 B_{ni}(t_1)x_i + B_{n0}(t), \;\;\; n=1,2,3,
\end{eqnarray}
where 
\begin{eqnarray}\label{BB}
&&
B_{ni}(t) =  a_{n1} A_{1i}(t_1) -  a_{n2} A_{2i}(t_1) + a_{n3}  A_{3i}(t_1),\;\;i=1,2,3,\;\; \\\nonumber
&&
B_{0}(t) =  a_{n1} A_{10}(t_1)- a_{n2} A_{20}(t_1)+ a_{n3} A_{30}(t_1) -\frac{a_{n3}}{2}.
\end{eqnarray}
The system of three equations (\ref{JJ2}) is uniquely solvable for $x_i$, $i=1,2,3$, if
\begin{eqnarray}
\label{B2}
\det\{B_{ni}(t_1): n,i=1,2,3\}\neq 0.
\end{eqnarray} 
Only in this case we are able to obtain the complete information about the initial state $\rho^S(0)$.
If $0<{\mbox{rank}}\,\{B_{ni}(t_1)\}<3$, then not all elements of the matrix $\rho^S(0)$ may be uniquely found (partial information transfer). And no information may be transfered if 
${\mbox{rank}}\,\{B_{ni}(t_1)\}=0$.
The system (\ref{B2}) is  equivalent to the  system (\ref{detA}).

Remark that coefficients  $B_{ni}(t_1)$ in the multi-channel communication system  
characterize the evolution operator $U(t)$, the  initial state $\tilde \rho(0)$ 
(through $A_{ij}$) and the  directions of the measured polarizations
 (through $a_{ni}$). { We may state that the proper choice of the  evolution operator and  the initial state provides the information transfer, while 
 the proper choice of  the directions of polarization measurements provides the  extraction of the transfered information.}

For instance, let the initial state of each channel is described by the following density matrix
\begin{eqnarray}
&&
\rho(0) =\rho^S(0)\otimes \tilde \rho(0),\;\;\;
\rho^S(0)\in H^S,\;\;\tilde \rho(0) \in H^{TL}\cup H^{R}.
\end{eqnarray}
Let, for instance,
\begin{eqnarray}\label{a}
a_{ni}=\delta_{ni}.
\end{eqnarray}
Then the system (\ref{BB}) defining coefficients $B_{nj}$ in the linear system (\ref{JJ2})
reads
\begin{eqnarray}\label{BBexp}
&&
B_{1i}(t_1) =   A_{1i}(t_1),\;\;
B_{10}(t_1) =  A_{10}(t_1),\;\;i=1,2,3, \\\nonumber
&&
B_{2i}(t_1) = -  A_{2i}(t_1) ,\;\;
B_{20}(t_1) = - A_{20}(t_1),\;\;i=1,2,3,\;\; \\\nonumber
&&
B_{3i}(t_1) =   A_{3i}(t_1),\;\; 
B_{30}(t_1) =  A_{30}(t_1) -\frac{1}{2}.
\end{eqnarray}
 Then Eq.(\ref{B2}) yields: $\det\{B_{ni}(t_1)\}=- \det\{A_{ni}(t_1)\}$, so that condition (\ref{B2}) coincides with   condition (\ref{detA}). 
Therefore the measurements of  polarizations in three mutually orthogonal directions $\mathbf{a}_n$ fixed by parameters (\ref{a}) 
may be used for obtaining the elements of  the  matrix $\rho^S(0)$ from the matrix $\rho^R(t_1)$. 

Note that the examples of three- and four-node chains considered above may be realized using the three channel scheme considered in this subsection.

\section{Conclusion}
\label{Section:conclusions}
Due to the quantum interactions, the information concentrated initially in a part of a quantum system (which is called the sender $S$)
becomes distributed  over the whole system. Thus, it may be completely extracted from another part of a quantum system (which is called the receiver $R$). Therefore, we may transfer  the information  from $S$ to $R$. The important property of such transfer is that it is not sensitive to the initial state of the system as well as  to the particular  type of quantum interactions. We arrange this transfer along the three- and four-node chain with nearest-neighbor interactions governed by the XY Hamiltonian. 
However there is no restrictions on the  length of the chain as  well as on the type of Hamiltonian, 
provided that  condition (\ref{detA}) is satisfied. We show that the elements of $\rho^S(0)$ may be extracted from the elements of $\rho^R(t_1)$ at some instant $t_1$ using the multi-channel communication system with the proper polarization measurements at the receivers of these channels.

Finally we  emphasize that the necessity to resolve the linear system (i.e. Eq.(\ref{A})) should not be considered as an essential disadvantage of the information transfer  approach in comparison with the perfect state transfer. This is especially valid   in the case of quantum computation algorithms. In fact, one has to remember, that  the algorithms for quantum computation must be able to handle the non-unitary transformations because  basic mathematical transformations  (such as  addition and multiplication) are  not  unitary ones \cite{NC}. Moreover,   the solution to the linear system (like the system (\ref{A}))  is one of the central problems of the mathematical calculations. Of cause,  the  algorithms of quantum computation must be able to  resolve such a  problem. 
{We also emphasize that the condition for the perfect state transfer is embedded into the condition for the complete information transfer. For this reason we conclude that the complete information transfer is much simpler for the realization in comparison with the perfect state transfer.}

Authors thank Professor E.B.Fel'dman for useful comments.
This work is supported by the Program of the Presidium of RAS No. 8 ''Development of methods of obtaining chemical compounds and creation of new materials''.


\begin{thebibliography}{99}


\bibitem{Bose}
S.Bose, Phys.Rev.Lett, {\bf 91} 207901 (2003)


\bibitem{CPW}
F.M.Cucchietti, H.M.Pastawski and D.A. Wisniacki, Phys. Rev. E {\bf 65}, 045206(R) (2002)


\bibitem{FBE}
E.B.Fel'dman, R.Br\"uschweiler and R.R.Ernst, Chem.Phys.Lett. {\bf 294}, 297 (1998)



\bibitem{CDEL}
 M.Christandl, N.Datta, A.Ekert and A.J.Landahl, Phys.Rev.Lett. {\bf 92}, 187902 (2004)

\bibitem{ACDE}
 C.Albanese, M.Christandl, N.Datta and A.Ekert, Phys.Rev.Lett. {\bf 93}, 230502 (2004)


\bibitem{KS}
 P.Karbach and J.Stolze, Phys.Rev.A {\bf 72}, 030301(R) (2005)


\bibitem{KF}
 E.I.Kuznetsova and E.B.Fel'dman, J.Exp.Theor.Phys. {\bf 102}, 882 (2006)

\bibitem{KZ}
 E.I.Kuznetsova and A.I.Zenchuk, Phys.Lett.A {\bf 372},  pp.6134-6140 (2008)

\bibitem{GKMT}
 G.Gualdi, V.Kostak, I.Marzoli and P.Tombesi, Phys.Rev. A {\bf 78}, 022325 (2008)


\bibitem{FPPK}
C.Di Franco, M.Paternostro, G.M.Palma and M.S.Kim, Phys.Rev.A {\bf 76}, 042316 (2007)

\bibitem{FPK}
 C.Di Franco, M.Paternostro and M.S.Kim, Phys.Rev.Lett. {\bf 101}, 230502 (2008)


\bibitem{BGB}
 D.Burgarth, V.Giovannetti and S.Bose, Phys.Rev.A {\bf 75}, 062327 (2007)


\bibitem{DZ}
S.I.Doronin, A.I.Zenchuk, 
 Phys.Rev.A {\bf 81}, 022321 (2010)  

\bibitem{BB}
A.Bayat, S.Bose, Phys.Rev.A {\bf 81},  012304 (2010)

\bibitem{YBB}
S.Yang, A.Bayat, S.Bose, Phys.Rev.A {\bf 84}, 020302(R) (2011)


\bibitem{WHDS}
L.Wang, J.-H.Huang, J.P.Dowling and S.-Y.Zhu, arXiv:1106.5097v2 [quant-ph]

\bibitem{JKMW}
D.F.V.James, P.G.Kwiat, W.J.Munro and A.G.White, 
Phys. Rev. A {\bf 64}, 052312 (2001) 


\bibitem{H}
Z. Hradil, Phys. Rev. A {\bf 55}, R1561 (1997)

\bibitem{BAPS}
 K. Banaszek, G.M. D'Ariano, M.G.A. Paris, and M.F. Sacchi,
Phys. Rev. A {\bf 61}, 010304 (1999)

\bibitem{HSBR}
Z. Hradil, J. Summhammer, G. Badurek, and H. Rauch, Phys.
Rev. A {\bf 62}, 014101 (2000)

\bibitem{RHJ}
J. \v{R}eh\'a\v{c}ek, Z. Hradil, and M. Je\v{z}ek, Phys. Rev. A {\bf 63}, 040303
(2001)



\bibitem{NC}
M.A.Nielsen and I.L.Chuang, Quantum Computation and Quantum Information, Cambridge
University Press, New York (2000)


\end{thebibliography}
\end{document}